\documentclass{Interspeech2024}

\interspeechcameraready 

\usepackage[utf8]{inputenc}
\usepackage[T1]{fontenc}
\usepackage{babel}
\usepackage{multirow} 
\usepackage{hhline} 
\usepackage{subcaption}
\usepackage{hyperref}
\usepackage{seqsplit}
\usepackage{ragged2e}
\usepackage{etoolbox}

\providecommand{\zd}[1]{\textcolor{black}{{#1}}}

\providecommand{\rr}[1]{\textcolor{black}{{#1}}}
\providecommand{\rr}[1]{\textcolor{black}{[{\bf #1}]}}

\newcounter{authorcount}
\makeatletter
\newcommand{\newname}[2][]{%
    \if\relax\detokenize{#1}\relax
        \def\author@affiliation{}%
    \else
        \def\author@affiliation{$^{#1}$}%
    \fi
    \ifnum\value{authorcount}=0
        \gdef\authorlist{#2\author@affiliation}%
    \else
        \xdef\authorlist{\unexpanded\expandafter{\authorlist}, #2\author@affiliation}%
    \fi
    \stepcounter{authorcount}
}
\makeatother

\title{GTR-Voice: Articulatory Phonetics Informed Controllable Expressive Speech Synthesis}

\newname[1]{Zehua Kcriss Li}
\newname[1]{Meiying Melissa Chen}
\newname[1]{Yi Zhong}
\newname[2]{Pinxin Liu}
\newname[1]{Zhiyao Duan}


\begin{document}
\raggedbottom

\address{
  \texorpdfstring{$^1$Department of Electrical and Computer Engineering, University of Rochester, Rochester, NY, USA
  $^2$Department of Computer Science, University of Rochester, Rochester, NY, USA}{1 Department of Electrical and Computer Engineering, University of Rochester, Rochester, NY, USA; 2 Department of Computer Science, University of Rochester, Rochester, NY, USA}
}

\email{\texorpdfstring{\{zehua.li, meiying.chen, zhiyao.duan\}@rochester.edu, yi.zhong@rutgers.edu, pliu23@u.rochester.edu}{zehua.li@rochester.edu, meiying.chen@rochester.edu, zhiyao.duan@rochester.edu,yi.zhong@rutgers.edu, pliu23@u.rochester.edu}}

\maketitle
 
\begin{abstract}
Expressive speech synthesis aims to generate speech that captures a wide range of para-linguistic features, including emotion and articulation, though current research primarily emphasizes emotional aspects over the nuanced articulatory features mastered by professional voice actors. 
Inspired by this, we explore expressive speech synthesis through the lens of articulatory phonetics. Specifically, we define a framework with three dimensions: Glottalization, Tenseness, and Resonance (GTR), to guide the synthesis at the voice production level. With this framework, we record a high-quality speech dataset named \textit{GTR-Voice}, featuring 20 Chinese sentences articulated by a professional voice actor across 125 distinct GTR combinations.
We verify the framework and GTR annotations through automatic classification and listening tests, and demonstrate precise controllability along the GTR dimensions on two fine-tuned expressive TTS models. 
We open-source the dataset and TTS models\footnote{Audio samples, visualization and code available at this link: \\ https://gtr-voice.com.}.
\end{abstract}
\section{Introduction}
Expressive speech synthesis aims to generate speech that captures a wide range of para-linguistic features~\cite{tan2021survey,review,eetts_paralinguistic}. A successful system could find broad applications in media production, education, and entertainment, creating a realistic and immersive experience for users. In recent years, deep-learning based speech synthesis methods have achieved high quality and naturalness \cite{tan2024naturalspeech, VC_chen, yang2023instructtts}. On expressiveness, state-of-the-art methods show good emotion rendering \cite{guo2023emodiff, zhu2024metts,emotts_finegrained} and style imitation \cite{li2024styletts, vyas2023audiobox}. However, compared to humans especially professional voice actors, their expressiveness and controllability are still very limited.      
There are many expressions that voice actors can do that these state-of-the-art methods cannot. An important reason, we argue, is that the scope of existing research is limited to certain aspects of speech expressiveness such as emotion and style, while many other aspects are simply not paid much attention by the research community.

According to the definition of expressive speech synthesis, speech expressiveness exists in various para-linguistic aspects. These aspects range from low-level aspects such as articulation and pronunciation to mid-level aspects such as speaking style, and to high-level aspects such as emotion and attitude. Existing research has been primarily focusing on mid- and high-level aspects, especially style \cite{style_contextaware, eetts_yizhong} and emotion \cite{Lei2022MsEmoTTS}, yet low-level aspects have not been paid much attention~\cite{deep_arti,Cao2017Integrating}. It is noted that low-level and high-level aspects can be independent of speech expression. For example, a professional voice actor is able to speak a sentence with diverse articulation methods but a particular emotion. Reversely, they can also speak with a particular articulation method but diverse emotions. The mid-level style aspect is related to the low-level articulation aspects, however, the former also includes prosodic features beyond the articulation and pronunciation of words. Furthermore, styles are categorical and not easy to manipulate (e.g., transitioning from one style to another, varying a style slightly) by humans or algorithms~\cite{voiceacting,voiceover}.

As expressive speech synthesis systems fall short of humans, it can be inspiring to learn what and how most capable humans, i.e., professional voice actors, can perform on speech expressiveness. In fact, they can manipulate many articulatory aspects to achieve the desired expression~\cite{actorvoice}.
For example, voice actors may be asked to change their expression by having a more breathy and softer tone, a more languid pronunciation, and a hint of nasality.
They understand how the three requirements correlate with specific vocal techniques to produce a voice that integrates these characteristics. Specifically, a more breathy and softer tone voice correlates with changes in the glottis configuration, a more languid pronunciation implies a variation in the muscle engagement of the articulators during articulating, and a nasal sound is achieved by manipulating the vocal tract shapes and configurations to alter the resonance cavities and acquire desired voice characteristics. In other words, there are a set of articulatory aspects that they can adjust to achieve the desired timbre and speaking style. In fact, throughout their curriculum, they learn to adjust many of these aspects independently and simultaneously.

Inspired by the capabilities of voice actors, in this paper, we investigate expressive speech synthesis from the perspective of articulatory phonetics. Specifically, we identify three fundamental dimensions of speech expression at the articulation level, namely Glottalization, Tenseness, and Resonance (GTR).
Under this framework, we designed and recorded a high-quality expressive speech dataset comprising 125 distinct GTR types of voice uttered by a single professional voice actor. The consistency of the dataset labels was evaluated through listener tests and automatic classification. Subsequently, we adapted two expressive Text-to-Speech (TTS) models (FastPitch~\cite{lancucki2021fastpitch} and StyleTTS~\cite{li2022styletts}) to investigate the feasibility of GTR control in expressive TTS. 
We conducted subjective evaluations on the GTR controllability and generated speech quality and naturalness. Results show that the adapted models exhibit fine control over each dimension of GTR in both Mandarin Chinese (same language) and English (cross-lingual) scenarios. 
\vspace{10pt}

\section{The GTR-Voice dataset}
The GTR-Voice dataset contains 3.6 hours of speech audio, comprising 2500 clips, with an average duration of 6 seconds each. All of the speech was recorded by a commercial-grade professional voice artist, a 30-year-old \zd{male} native speaker of Mandarin Chinese. The scripts consist of 20 utterances drawn from the Global TIMIT Mandarin Chinese pool~\cite{ctimit}. These utterances are part of a larger set of 120 sentences selected from the Chinese Gigaword Fifth Edition~\cite{gigaword}, which is recognized for its broad phonetic coverage. The GTR dataset is labeled with 5 Glottalization and 5 Tenseness labels, and 7 Resonance labels. Note that the label 0-Whisper in Glottalization corresponds directly to the label 0-Whisper in Resonance. This means that when Glottalization is labeled as 0-Whisper, Resonance should also be labeled as 0-Whisper. As a result, there are 125 distinct GTR types in total. The audio files are in mono WAV format, with a sampling rate of 48 kHz and a bit depth of 24 bits. 

\vspace{-0.59em} 

\subsection{Articulatory phonetics inspired dimensions}
{Articulatory phonetics provides a systematic framework for classifying the pronunciation attributes of phonemes in different languages~\cite{gick2013articulatory}. In this work, we apply  principles from articulatory phonetics and related speech and vocal production research~\cite{titze1998principles,voice_source}, and use their definitions from research to mark utterance-level articulation traits. We identify three fundamental dimensions of articulation: Glottolization~\cite{voice_quality_glottalization}, Tenseness~\cite{tenseness}, and Resonance~\cite{resonance_analysis}.}


\textbf{Glottalization} refers to the control of air flow due to the tension of the glottis (i.e., throat).
During the articulation of a vowel, 
varying degrees of openness of vocal cords lead to different phonation types~\cite{phonation_types_labels,phonation_types_in_speech_singing,phonation_cross_lin}. 
We use a five point scale (0-4) to label the phonation types: 

\begin{itemize}

\item\textbf{0 - Whisper \zd{Voice}:}
Air passes through the glottis without causing vocal cords vibration, resulting in sounds without a vocalic tone~\cite{whisper}.
\item\textbf{1 - Slack Voice:}
Vocal cords are loose and the airflow is average, resulting in sounds somewhat dull and lacking power, sometimes used to convey a laid-back, carefree mood.
\item\textbf{2 - Modal Voice:}
The most frequently used phonation type, where the vocal cords vibrate to form consistent and regular sound waves, making natural and pleasant voices.
\item\textbf{3 - Stiff Voice:}
Firm, frequently vibrating vocal cords yield a sharp, shrill voice. Such a tone is commonly used in voice-overs for tough characters or suspenseful scenarios.
\item\textbf{4 - Creaky Voice:}
Tight vocal cords with barely any airflow likened to an old rusty hinge and harsh sounding. More vocal force leads to harsher speech. It is usually found in strong emotional portrayals or cartoon character dubbing.
\end{itemize}

\textbf{Tenseness} measures the pronunciation quality and the tension of articulation muscles~\cite{tenseness,tense_lax}. Generally speaking, tense vowels in pronunciation involve tension in the tip and root of the tongue, while lax vowels are the opposite. We use a five point scale (1-5) to denote from laxest to the tensest pronunciation for both consonants and vowels:
\begin{itemize}
    \item \textbf{1 - Laxest:} 
        Minimal articulatory muscle use blurs vowel and consonant clarity, resembling a yawn. Used in dubbing to portray careless, tired, or low-energy characters.
    \item \textbf{2 - Slightly Lax:} 
        \zd{Engagement of articulators} slightly rises, maintaining a relaxed tone. Phonemes are clearer yet lax, sounding naturally relaxed, fitting for lazy-style ads or documentaries.
    \item \textbf{3 - Moderate:}
        Articulators are moderately engaged, balancing between tension and laxness, ideal for standard speech with clear, balanced vowels and consonants, ensuring high intelligibility.
    \item \textbf{4 - Slightly Tense:} 
        \zd{Tension of articulators} increases for vigorous, clear pronunciation, often used to highlight specific words or emotions.
    \item \textbf{5 - Tensest:} 
        Articulator muscles exert much force, akin to gritting one's teeth, intensifying speech timbre and clarity, ideal for expressing anger, excitement, or tension.
\end{itemize}

\textbf{Resonance}  refers to the integration of articulatory phonetics with vocal register insights~\cite{Hollien1974On}. We use seven labels, including two basic voices and four mixed voices~\cite{mixed_voice,chest_head,chest_head_falsatto_voice,mixed_chest_falsatto} including the nasality~\cite{Amino2014Nasality}, along with the whisper voice:
\begin{itemize}
    \item \textbf{0 - Whisper Voice:}
        Air passes through the glottis without causing vocal cords vibration, resulting in sounds without a vocalic tone~\cite{whisper}.

    \item \textbf{1 - Chest Voice:} 
        Characterized by a sensation of vibration within the chest cavity, generating a powerful and resonant sound quality. Frequently used in voice acting to show strong, authoritative characters.

    \item \textbf{2 - Head Voice:} 
        Produces a vibrant, light sound in the head, giving a sense of joy and playfulness, often tied to a higher pitch, aligning with falsetto, which can be viewed as a specific form of head voice, characterized by its airy texture and elevated pitch range.
    \item \textbf{3 - Chest-Nasal Mix:} 
        Uses chest resonance and nasal modulation for a warm and soft sound, often signaling sincerity or cosiness.
    \item \textbf{4 - Chest-Head Mix:}
        Merges chest voice power with head voice clarity for a versatile, strong, high-pitched tone. Belting skill in singing is based on this type of vocalization method.
    \item \textbf{5 - Head-Nasal Mix:}
        Blends head voice lightness with nasal resonance for a vibrant, dramatic sound, perfect for scenario like mischievous prank.
    \item \textbf{6 - Chest-Head-Nasal Mix:} 
        A kind of mixed voice that sounds like it comes from a naive and cute cartoon character, often used as voice-over for humorous scenes.
\end{itemize}


\subsection{Recording process}

To produce voices that follow the designed dataset labels, the voice actor, who has outstanding performative voice rendering techniques, comprehensively cogitated and evaluated the voice recording process to ensure that the produced voices adhere to the GTR labels. 

Specifically, the voice actor leverages self-evaluation for training, based on the dataset label guidance and his skills in articulation modulation. They first start with a GTR combination that is easier \zd{to voice} than others \zd{are}, such as G=2, T=3, R=1, then record the clip and explore the variants based on this clip as a reference. The voice actor then continues the process by varying a specific dimension. For example, to record the variations for Glottalization, the voice actor might move from G2T3R1 to G3T3R1, while keeping the other dimensions constant. The voice actor continues this process until all sample voice clips for the Glottalization label \zd{variations} are recorded. This strategy is utilized to explore all combinations of dataset labels. 




Data was recorded in a professional recording studio with minimum background noise and reverberation, using the Neumann M249 microphone, the NEVE 4081 pre-amplifier, and the Logic Pro X software. This setup is considered a professional setup in commercial audio engineering practices.

\begin{figure*}[t]
    \centering    \includegraphics[width=1.\linewidth]{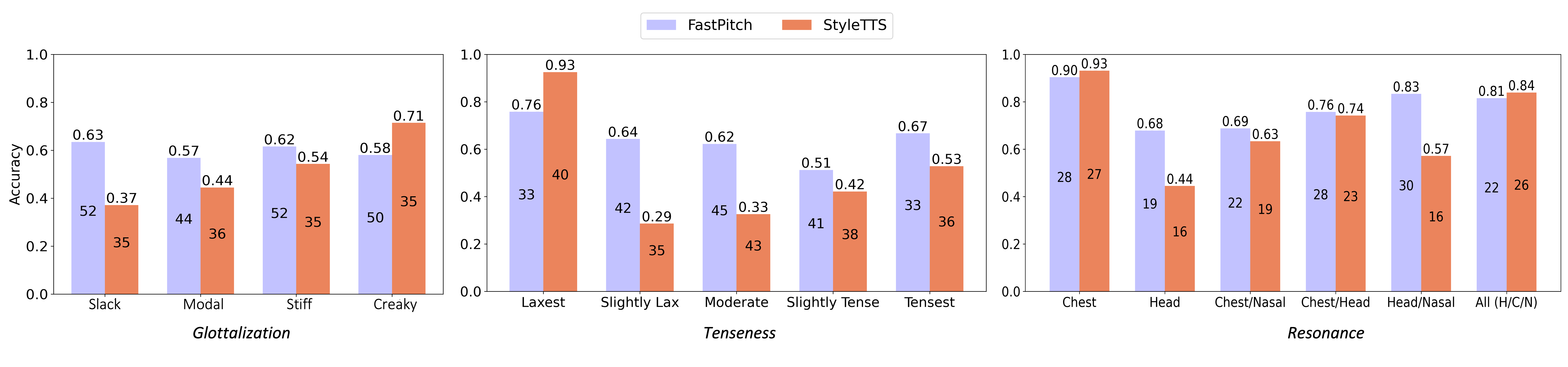}
    \vspace{-0.8cm}
    \caption{\zd{Average classification accuracy across subjects and utterances of the subjective evaluation of GTR controllability for FastPitch and StyleTTS. Sub-figures from left to right show results of GTR variations along different dimensions around the value of the reference utterance. 
    We do not include the Whisper voices (G=0 \& R=0) as only Tenseness can vary, making it unfair to compare with other GTR variations, based on the evaluation process we conduct.}}
    \label{fig: subjective_compare}
    \vspace{-0.5cm}
\end{figure*}


\subsection{Label Verification}
To ensure the reliability of the labels, we conduct both human evaluation and automatic classification. Both results confirmed the consistency of the labels of our dataset.

\textbf{Human Evaluation:} Three speech processing researchers were asked to label 50 random samples from the dataset. Before they started, they were first trained by the voice actor who recorded the dataset for 1 hour. During the labeling process, they had access to recordings of all GTR combinations for one sentence. 
The average correlation across the three participants between the ground-truth labels and their predictions are 0.75 and 0.80 for G and T dimensions, respectively. Since the R dimension is not ordered, we use the classification accuracy, and the average accuracy on R is 0.78.

\textbf{Automatic Classification:} We trained three SVM classifiers, one on each dimension using the following features extracted by librosa\footnote{\url{https://github.com/librosa/librosa}}: MFCC, mel spectrogram, spectral contrast, spectral centroid, spectral bandwidth, spectral roll-off, min of spectral roll-off, F0, voiced flag, voiced probs, zero crossing rate, and flatness. We also tried training two regressors on G and T dimensions but obtained significantly worse results compared with the classifiers. The accuracies of the three classifiers \zd{corresponding to} the GTR dimensions are 0.85, 0.83, and 0.94, respectively.



\section{GTR controllable speech synthesis}
One goal of this work is to test the feasibility of training expressive speech synthesis models with GTR controls on the proposed GTR-Voice dataset. To do so, we adapt two text-to-speech (TTS) models to incorporate GTR labels as control conditioning. 

\subsection{FastPitch}
FastPitch~\cite{lancucki2021fastpitch} is a feedforward Transformer~\cite{vaswani2017attention} TTS model, which includes a bidirectional Transformer encoder, plus pitch and duration predictors. Initially, the signal is encoded by the first 6 Transformer Encoder blocks, then augmented with pitch and upsampled to frame level by the length regulator. Afterward, it passes through another 6 Transformer Decoder blocks to refine and create a mel spectrogram. We achieve GTR control by adding conditions to the Encoder output of FastPitch. These conditions are obtained through three separate trainable embedding layers, each dedicated to embedding one dimension of the GTR labels.

We utilize the implementation of the official FastPitch repo\footnote{\href{https://github.com/NVIDIA/DeepLearningExamples/tree/master/PyTorch/SpeechSynthesis/FastPitch}{\texttt{\seqsplit{https://github.com/NVIDIA/DeepLearningExamples/tree/master/PyTorch/SpeechSynthesis/FastPitch}}}}. We first pre-trained it with \rr{a Chinese dataset AI-SHELL3}~\cite{shi2020aishell} with the default configuration for 80 epochs. And then fine-tune the entire network conditioned on GTR label embeddings on our proposed dataset. The dimension of these three embedding layers is 384, aligned with the dimension of the encoder output. We fine-tune the model with a start learning rate of 0.01 and inverse square root learning rate decay for 3000 epochs. We use the Adam optimizer with a weight decay of 1e-6 following the open-source setup. During both training, we used a batch size of 32 with a single RTX 3090 GPU.

\subsection{StyleTTS}
StyleTTS~\cite{li2022styletts} is a style-based parallel TTS model that synthesizes speech with naturalistic prosody and emotional tones by learning from reference speech utterances. It employs a style encoder to extract style embeddings from the references. The embeddings are then used as conditioning in the synthesis process, generating speech that closely matches the prosodic and emotional characteristics of the reference utterances.

StyleTTS' training process consists of two stages: Mel-spectrogram reconstruction and duration/prosody prediction, with the style encoder being trained unsupervisedly across both stages. The style encoder takes mel-spectrogram as input to produce a fixed-dimension style embedding that captures the unique prosodic and emotional attributes of the input.

To adapt StyleTTS for use with GTR labels instead of mel-spectrogram as input, we leveraged a pre-trained style encoder from an existing model. We paired each utterance in our dataset with its GTR label, and then input the utterance into the pre-trained style encoder to generate a style embedding. We developed a small neural network called GTR embedder to predict these extracted style embeddings from the corresponding GTR labels. After training, the original style encoder was replaced with the GTR embedder, while the rest of the pre-trained components remained unchanged.

We utilized the official implementations and model checkpoints provided by its authors\footnote{\url{https://github.com/yl4579/StyleTTS}}. The model was pre-trained on the multi-speaker English dataset LibriTTS~\cite{zen2019libritts} train-clean-460 subset for 200 epochs across both stages. The GTR embedder consists of one embedding layer and two fully connected layers with ReLU activation. This embedder is exclusively trained on our proposed dataset, using a single RTX 3090 GPU for 500 epochs. The remaining network weights for StyleTTS are derived from the pre-trained checkpoints.



\begin{table}[t]
\small
\centering
\setlength{\tabcolsep}{3mm}

\begin{tabular}{c|cc}
\hline
Model & Quality$\uparrow$ (1-5) & Naturalness$\uparrow$ (1-5) \\
\hline
\multirow{2}{*}{FastPitch} & \multirow{2}{*}{3.05 $\pm$ 0.05} & \multirow{2}{*}{3.14 $\pm$ 0.11} \\
&&\\
\hline
\multirow{2}{*}{StyleTTS} & \multirow{2}{*}{4.21 $\pm$ 0.14} & \multirow{2}{*}{4.16 $\pm$ 0.12} \\
&&\\
\hline
\end{tabular}
\vspace{0.2cm}
\caption{\zd{Average and standard deviation of Mean Opinion Score (MOS) across all utterances and subjects for each model in the subjective evaluation.} Each score is on a 5-point scale (1-Bad, 2-Poor, 3-Fair, 4-Good, 5-Excellent). Higher scores indicate better performance.}
\label{table_user_study} 
\vspace{-0.5cm}
\end{table}

\section{\zd{Experiments}}

%

\subsection{Evaluation Setup}
A user study involving 60 general public participants was conducted to assess the GTR controllability, speech quality, and naturalness of generated speech using FastPitch and StyleTTS. 
Specifically, we shuffled our dataset and designed 40 evaluation webpages within our study, there are 20 for Chinese (FastPitch) and the others for English (StyleTTS), each containing 18 or 19 trials. The scripts for generated audio in Chinese evaluation are 20 utterances randomly sampled from AISHELL-3, while in English evaluation are 20 utterances randomly sampled from LibriTTS.
For each trial, we presented one bonafide audio clip, which are from GTR dataset, along with three audio clips generated from the model. The generated clips were obtained by fixing two of the GTR dimensions while varying the third. For example, to evaluate the model's controllability on the R dimension, the variation \zd{was} toward the close levels for R while making sure G and T are consistent with the reference bonafide audio. We \zd{asked} the participants to respond to the following statements: \textit{``Which speech sample (Sample 1, Sample 2, Sample 3) sounds most similar to the reference?"}. After users \zd{chose} the most similar audio generated by the model, we additionally \zd{asked} them to rate the Mean Opinion Score (MOS) of speech quality and speech naturalness separately.


\subsection{Results \zd{and Discussions}}
Tab.~\ref{table_user_study} presents the \zd{average and standard deviation values of} Mean Opinion Score (MOS) results of speech quality and naturalness \zd{across all test utterances and subjects} in the user study. The scores, ranging from 1 (Bad) to 5 (Excellent), \zd{suggest descent performance of both the FastPitch and StyleTTS models with both average quality and naturalness scores above 3, laying the foundation for our following controllability experiments.} 



Fig.~\ref{fig: subjective_compare} shows the accuracy of choosing the most similar generated audio during the test. Overall, participants achieve high accuracy across all three dimensions, demonstrating the feasibility of GTR control. Specifically, FastPitch achieves an accuracy of 67.07\%, and StyleTTS achieves 57.14\%. The highest performance is observed for the R dimension (78.01\% for FastPitch and 68.64\% for StyleTTS), while the lowest is for the G dimension (60.10\% for FastPitch and 51.77\% for StyleTTS). For context, random guess accuracy is 33.33\%, so all reported accuracies are significantly above chance, indicating controllability.

Interestingly, this finding is in contrast to the higher speech quality and naturalness scores achieved by StyleTTS in the MOS evaluation. We hypothesize that this difference may be attributed to the language-specific pretraining data and the different training strategies employed for each model. During the fine-tuning of FastPitch, the entire network was updated, which resulted in better learning of various GTR characteristics compared to StyleTTS. However, the sound quality was not well-maintained. StyleTTS was pre-trained \zd{on English data} and only the style embedder \zd{was} finetuned on the GTR dataset. While this strategy helps maintain the synthesis quality, it may hinder the model's ability to effectively adapt to the specific characteristics of the GTR dataset through transfer learning. 

\setlength{\RaggedRightParindent}{\parindent}
\RaggedRight
\sloppy

\indent Regarding the Glottalization dimension, participants show consistent performance across all four categories for FastPitch. However, the performance for StyleTTS varies considerably, with the lowest accuracy for Slack Voice and the highest for Creaky Voice. Moreover, a Chi-square test ($p=0.026$) indicates that participants' accuracy for StyleTTS differs significantly across the G dimension categories. A post-hoc analysis with Bonferroni correction reveals that the primary difference stems from the performance gap between Slack and Creaky Voices $(p < 0.008)$.
\justifying


For the Tenseness dimension, participants achieve the highest accuracy for the Laxest category. We posit that the Laxest voice exhibits the most distinct acoustic properties compared to the other categories, leading to better perceptual discrimination. Except Laxest, for other categories of Tenseness, the accuracy for audio from StyleTTS is significantly lower than that from FastPitch $(p < 0.01)$.

In the Resonance dimension, Chest Voice yields the highest predictability. We hypothesize that this is because Chest Voice is an in-domain category for the Expressive-TTS pre-trained model, while the other categories, particularly Head Voice and Head/Nasal Voice, occur less frequently. This is evident from Fig.~\ref{fig: subjective_compare}, which shows that StyleTTS predictions for these two categories have the lowest performance ($p < 0.001$, chi-square test). Post-hoc analyses revealed significant differences between Chest Voice and Head Voice $(p < 0.001)$, Head Voice and all three other categories $(p < 0.001)$, and Chest Voice and Head/Nasal Voice $(p < 0.01)$.

\section{Conclusions and future work}
This paper presented the GTR-Voice dataset and an innovative \zd{perspective toward} expressive speech synthesis by drawing inspirations from speech production research, particularly articulatory phonetics. Through the investigation of Glottalization, Tenseness, and Resonance dimensions, we demonstrated the potential for achieving controllable expressive TTS at \zd{the articulatory phonetics level}. 
However, the variations \zd{of controllability} across different GTR dimensions highlight the need for further research. 
For future work, we \zd{plan to improve the quality and GTR controllability of expressive TTS systems, and explore multi-speaker and cross-lingual TTS scenarios.} 

\section{Acknowledgements}
This work is supported in part by a New York State Center of Excellence in Data Science award and synergistic activities funded by NSF grant DGE-1922591.

\bibliographystyle{IEEEtran}

\bibliography{mybib}

\end{document}